\documentclass[twocolumn,floatfix,superscriptaddress,a4paper,showpacs,showkeys,nofootinbib,reprint,prc,noeprint]{revtex4-1}
\usepackage{epsfig}
\usepackage{latexsym}
\usepackage[colorlinks=true,linktocpage=true,linkcolor=blue,citecolor=blue,allcolors=blue]{hyperref}
\usepackage{url}
\usepackage[utf8]{inputenc}
\usepackage{enumerate}
\usepackage{color}
\usepackage{xcolor}

\usepackage{xfrac}

\usepackage{amsmath}
\usepackage{amssymb}

\usepackage[english]{babel}
\usepackage{hyperref}
\usepackage{url}
\usepackage{needspace}
\usepackage{float}
\hypersetup{
   colorlinks=true, 
  linktoc=all,     
  linkcolor=blue,  
}

\newcommand{\eq}[1]{\begin{align} #1 \end{align}}

\begin{document}

 \title{Charmed nuclei and exotic charmed meson production \\ at CBM@FAIR and ALICE@LHC}

\author{Thanaporn Chimruang}
\affiliation{School of Physics, and Center of Excellence in High Energy Physics and Astrophysics, Suranaree University of Technology, Nakhon Ratchasima, 30000, Thailand}
\affiliation{GSI Helmholtzzentrum f\"ur Schwerionenforschung GmbH, Planckstr. 1, D-64291 Darmstadt, Germany}

\author{Christoph Herold}
\email{herold@g.sut.ac.th}
\author{Ayut Limphirat}
\email{ayut@g.sut.ac.th}
\author{Yupeng Yan}
\affiliation{School of Physics, and Center of Excellence in High Energy Physics and Astrophysics, Suranaree University of Technology, Nakhon Ratchasima, 30000, Thailand}

\author{Jan Steinheimer}
\affiliation{GSI Helmholtzzentrum f\"ur Schwerionenforschung GmbH, Planckstr. 1, D-64291 Darmstadt, Germany}
\affiliation{Frankfurt Institute for Advanced Studies, Ruth-Moufang-Str. 1,  60438 Frankfurt am Main, Germany}

\author{Benjamin D\"{o}nigus}
\affiliation{Institut f\"{u}r Kernphysik, Goethe-Universit\"{a}t Frankfurt, Max-von-Laue-Str. 1, D-60438 Frankfurt am Main, Germany}

\author{Tom Reichert}
\affiliation{Theoretical Physics Department, CERN, 1211 Geneva 23, Switzerland}

\author{Volodymyr Vovchenko}
\affiliation{Department of Physics, University of Houston, Houston, TX 77204, USA }

\author{Marcus Bleicher}
\affiliation{Institut f\"{u}r Theoretische Physik, Goethe-Universit\"{a}t Frankfurt, Max-von-Laue-Str. 1, D-60438 Frankfurt am Main, Germany}
\affiliation{Helmholtz Research Academy Hesse for FAIR (HFHF), GSI Helmholtzzentrum f\"ur Schwerionenforschung GmbH, Campus Frankfurt, Max-von-Laue-Str. 12, 60438 Frankfurt am Main, Germany}

\date{\today}


\begin{abstract}
We make predictions for the expected multiplicities of exotic charmed hadrons and charmed nuclei in Au+Au collisions at SIS100 and LHC beam energies, using input on light hadron and charm production from the UrQMD transport model, and applying the Thermal-FIST model. 
We demonstrate that the CBM experiment has the capability to explore these states with production rates of
one per 3 seconds for $\chi_{c0}(1P)$ and $\chi_{c1}(1P)$, and one every 3 minutes for $X(3872)$ at the expected data taking rates. 
Due to the higher baryon density at CBM compared to the LHC, charmed nuclei, if they exist, will be equally abundant at CBM as at the LHC, even though the total charm production at CBM is much lower.
\end{abstract}

\maketitle

\section{Introduction}

Charmed hadrons are an important tool to test our understanding of the strong interaction \cite{Matsui:1986dk,Harris:2023tti}. Especially the study of so-called exotic states of charmed mesons can give new insights and dedicated experimental programs exist, e.g., at BESIII \cite{Wang:2024hxp} and LHCb \cite{Palano:2015fta}. A downside of such programs is the rather small production probability of charmed hadrons due to their large mass, even in high energy $e^+e^-$ or hadron-hadron collisions. Another interesting topic relating to charmed physics is the possible existence of charmed nuclei \cite{Starkov:1986ye,Tsushima:2003dd,Steinheimer:2016jjk}, i.e., atomic nuclei containing at least one charmed hadron. Requiring the coalescence of charmed baryons with at least one other baryon makes such states difficult to produce.

High-energy collisions of heavy nuclei, on the other hand, offer a unique opportunity for producing exotic charmed hadrons as well as charmed nuclei. Here, charm quarks are produced in the early stage of the reaction and their number is conserved throughout the evolution of the fireball. Their distribution into various hadron species is then determined at the chemical freeze-out \cite{Braun-Munzinger:2009dzl,Andronic:2021erx}. In addition, due to the large number of baryons involved in the collisions, the probability to produce light nuclei in the collision zone is large enough to allow for the formation of charmed nuclei \cite{Schaffner-Bielich:1998ogl,Steinheimer:2016jjk}.

In this work, we provide estimates of the production cross sections of several exotic charmed hadrons as well as charmed nuclei in heavy-ion reactions at different beam energies at the future SIS100 accelerator and the existing LHC. Using the multiplicities of light hadrons and the most common charmed hadrons as input, we can use the assumption of statistical hadron production \cite{Andronic:2021erx} to estimate the yields of exotic charmed hadrons and charmed nuclei with the Thermal-FIST model \cite{Vovchenko:2019pjl}. Within the statistical hadronization model (SHM)~\cite{Andronic:2017pug}, the system is described as a thermalized hadron gas, and the partition function encodes the accessible states. This framework allows to predict the production yields and relative abundances of charmed hadrons~\cite{Andronic:2021erx,Andronic:2017pug}, including open charm mesons, charmed baryons, hidden charm states, and possible exotic configurations. 

The light and charmed hadron multiplicities are estimated for the SIS100 beam energy range using the microscopic transport model UrQMD \cite{Bass:1998ca,Bleicher:1999xi,Bleicher:2022kcu} (for recent charm studies at FAIR energies within UrQMD see e.g. \cite{Steinheimer:2016jjk,Reichert:2025iwz}). At LHC energies we can rely on available data from the ALICE experiment as input.

Such predictions are highly relevant for upcoming experiments at FAIR and GSI as well as at the LHC, where the observation of charm and exotic bound states would provide valuable information on QCD in extreme conditions \cite{Messchendorp:2025men}.

\section{Methodology}

\subsection{Thermal hadron yields}

The statistical hadronization model has been used in the past to describe the multiplicities of hadrons produced in Au+Au, p+Au as well as p+p collisions over a wide range of beam energies \cite{Becattini:1997rv,Cleymans:1999st,Andronic:2005yp,Andronic:2017pug}. It is based on the assumption of chemical equilibrium where hadrons are produced according to thermal weights at the chemical freeze-out stage of the collision.\\

Due to the large mass of the charm quark, charm–anticharm pairs in nuclear collisions are predominantly created in initial hard partonic scatterings at high energies or by hadronic multi-step process at lower energies and their total number remains conserved during the subsequent evolution of the system. At beam energies relevant for FAIR and even at moderate LHC energies, the average number of charm pairs per event is significantly smaller than unity. As a consequence, charm production cannot be treated within the grand canonical ensemble but requires an exact, canonical implementation of charm conservation.\\

In a general canonical statistical hadronization approach, the yields of charmed hadrons, in equilibrium, would be suppressed relative to their grand-canonical expectation. This suppression originates from exact conservation of the charm quantum number within a finite correlation volume ($V_c$) and becomes increasingly important for small charm multiplicities (similar to the effect observed for strangeness in small systems and beam energies close to the strangeness threshold \cite{Becattini:1997rv,Hamieh:2000tk}).\\

The canonical formulation of the statistical hadronization model is based on
the exact conservation of the relevant quantum numbers within a finite
correlation volume. 
In this approach, the partition function is constructed by projecting the grand-canonical ensemble onto fixed values of the conserved charges.
For a hadron resonance gas with conserved baryon number $B$, electric charge $Q$, strangeness $S$, and charm $C$, the canonical partition function at a given temperature $T$ and correlation volume $V_c$ reads~\cite{Becattini:1995if,Becattini:1997rv,Vovchenko:2019kes}:
\eq{\label{eq:Z}
& \mathcal{Z}(B,Q,S,C)  =
\int \limits_{-\pi}^{\pi}
  \frac{d \phi_B}{2\pi}
 \int \limits_{-\pi}^{\pi}
  \frac{d \phi_Q}{2\pi}
  \int \limits_{-\pi}^{\pi}
  \frac{d \phi_S}{2\pi}
    \int \limits_{-\pi}^{\pi}
  \frac{d \phi_C}{2\pi} \nonumber \\
  & \quad \times
  e^{-i \, (B \phi_B + Q \phi_Q + S \phi_S + C \phi_C)} \nonumber \\
  & \quad \times e^{\left[\sum_{j} \sum_{n=1}^{\infty} z_{j,n} \, e^{i \, n \, (B_j \phi_B + Q_j \phi_Q + S_j \phi_S + C_j \phi_C)}\right]}.
}
with
\begin{equation}
z_{j,n} = (\mp 1)^{n+1} \gamma_{q}^{|q_i|} \gamma_{S}^{|S_i|} \gamma_{C}^{|C_i|} \frac{d_j V_c}{2 \pi^2} \frac{T m^2}{n^2} K_2 \left( n \frac{m}{T}\right)
\end{equation}

Canonical corrections simplify considerably when the canonical treatment is applied to charm conservation only (in this case, the baryon, strangeness, and electric charge densities are regulated by the corresponding chemical potentials).
In the absence of multi-charm states, as is the case here, and applying the Maxwell-Boltzmann approximation ($n = 1$),
\eq{\label{eq:ncharm}
n_i^{C, \rm ce} = n_i^{C, \rm gce} \, \frac{I_1(2 \sqrt{S_{+1}^C S_{-1}^C})}{I_0(2 \sqrt{S_{+1}^C S_{-1}^C})} \, \sqrt{\frac{S_{\mp 1}^C}{S_{\pm 1}^C}}, \quad \text{for } C_i = \pm 1.
}
Here
\eq{
S_{\pm 1}^C = V_c \sum_{i \in C_i = \pm 1}  n_i^{C, \rm gce},
}
where $n_i^{C, \rm gce}$ is the grand-canonical density
\eq{
n_i^{C, \rm gce} = \gamma_{q}^{|q_i|} \gamma_{S}^{|S_i|} \gamma_{C}^{|C_i|} \frac{d_i T m^2}{2 \pi^2} K_2 \left( \frac{m_i}{T}\right) \, e^{\frac{\mu_B B_i + \mu_Q Q_i + \mu_S S_i}{T}}.
}

Using Thermal-FIST we verified that the canonical corrections for charges other than charm are negligible for Au+Au collisions studied here, and that Eq.~\eqref{eq:ncharm} provides an accurate description of charm yields.
For this reason, we treat other quantum numbers within a grand canonical partition function and fix the relevant temperature and chemical potentials using the standard chemical freeze-out systematics.

A unique feature about charm production in nuclear collisions is that, even though their multiplicity is very small, the charmed quark pairs are mainly produced in the early stage of the collision, and the total abundance of charm+anti-charm quarks is larger than what one would expect even for a grand canonical ensemble. Therefore, one assumes that charmed hadrons are only in partial equilibrium. A charm fugacity $\gamma_c$, which tends to be larger than unity, is introduced in addition the charm canonical radius $R_c$ ($V_c=\frac{4}{3}\pi R_c^3$) that only influences the relative abundances of the charmed hadrons.

\subsection{Model inputs: Open and hidden charm yields}

To make predictions for the statistical distribution of the charm quarks among the produced hadrons, the two additional inputs (in addition to temperature and the baryon-chemical potential) need to be constrained: the charm suppression/enhancement factor $\gamma_c$  and the charm canonical radius $R_c$. Typically one uses the total charm yield, and a hidden charm meson multiplicity as input.

Practically, there are two different ways to obtain these inputs: At higher energies, like currently explored at the LHC, one conveniently uses the measured multiplicities of light and charmed hadrons as input for the thermal parameter fits. This is possible for experiments like ALICE where an extensive set of measurements exists. Unfortunately, at lower energies and for reactions that have not yet been measured (like in the present case for FAIR energies), we cannot rely on existing data. 

Consequently, we use the available ALICE data \cite{ALICE:2021kfc,ALICE:2021rxa,ALICE:2023gco,ALICE:2021bib} for our predictions at LHC beam energies on charm production in Pb+Pb collisions at 5.02 TeV as input for the thermal fits. In the FAIR energy range, as a replacement for the yet unavailable experimental data, we use estimates based on the UrQMD model. The UrQMD transport model \cite{Bass:1998ca,Bleicher:1999xi,Bleicher:2022kcu} has been used successfully in the past to describe hadron multiplicities in hadron and nuclear reactions over a wide range of beam energies. It is based on hadronic degrees of freedom and the excitation and fragmentation of color strings at higher energies. In the latest version, the implemented hadron spectrum includes (strange and non-strange) meson and baryon resonances up to masses of 4 GeV. The hadrons are propagated via the Hamilton equations of motion, if nuclear potentials are used. Without potentials, the hadrons follow straight line trajectories (cascade mode). Typically, the collision of two nuclei consists of three stages: I) initialization of the colliding nuclei, II) particle production, decay, and subsequent rescattering, followed by III) the freeze-out/decoupling stages, where inelastic and finally all interactions cease.

A recent study has already shown that the chemical freeze out parameters obtained from the UrQMD evolution, when including a realistic equation of state, lie close to a global chemical freeze out line that is compatible with the thermal model parameters \cite{Inghirami:2019muf,Reichert:2020yhx}. It is therefore reasonable to use the temperature and chemical potentials from a fit to UrQMD as input for our predictions of expected hadron yields in the SIS100 beam energy range. 

    \begin{figure}[t]
        \centering
        \includegraphics[width=0.5\textwidth]{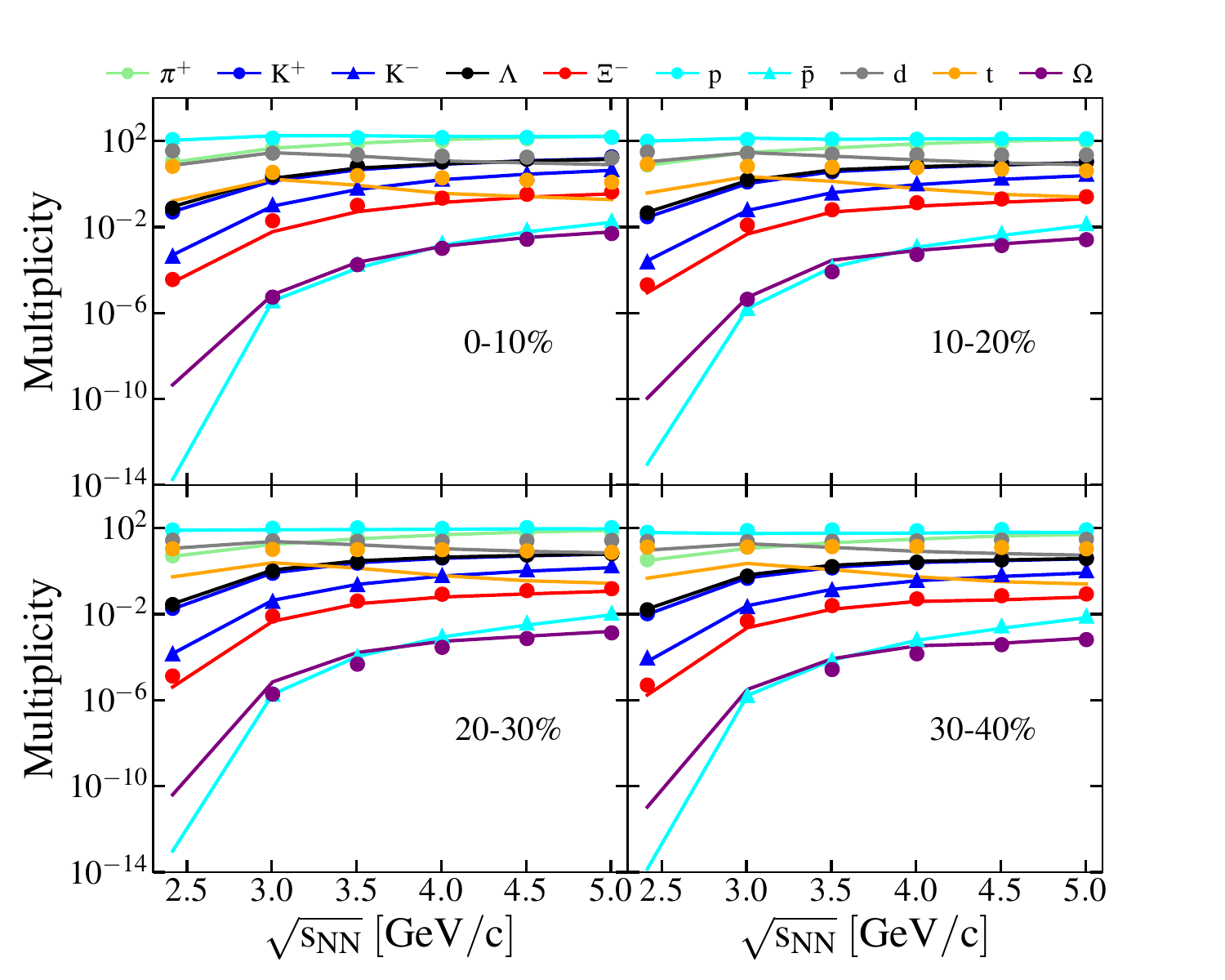}
        \caption{Input (symbols) and fitted (lines) light hadron multiplicities from the UrQMD simulations of Au+Au collisions at different centrality selections. For the fits only hadron multiplicities without light nuclei where used. }
        \label{fig:1}
    \end{figure}

To constrain the inputs of charmed hadron production $\gamma_c$ and $R_c$ needed for the statistical analysis, we use the open charmed hadron multiplicity from $\Lambda_c$ to fix $\gamma_c$ and the hidden charm hadron multiplicity from $J/\Psi$ to fix $R_c$. If data from FAIR become available, these estimates could be replaced by actually measured yields.

In a previous work, we have discussed how the production of these charmed hadrons at FAIR energies can be included in the UrQMD simulations. To calculate charm production near and even below threshold, it was implemented via the formation and decays of heavy resonances (in analogy to $\phi$ production), with parameters fitted to elementary cross sections measured by experiment in p+p interactions slightly above threshold \cite{Steinheimer:2016jjk}. While this approach is sufficient to predict the most common charm states, it is not detailed enough to determine the production of more exotic charmed states due to the lack of known branching ratios. That is why we have to rely on the statistical assumption to redistribute the charm quarks.

\section{Fit results - input and bulk}

    \begin{figure}[t!]
        \centering
        \includegraphics[width=0.5\textwidth]{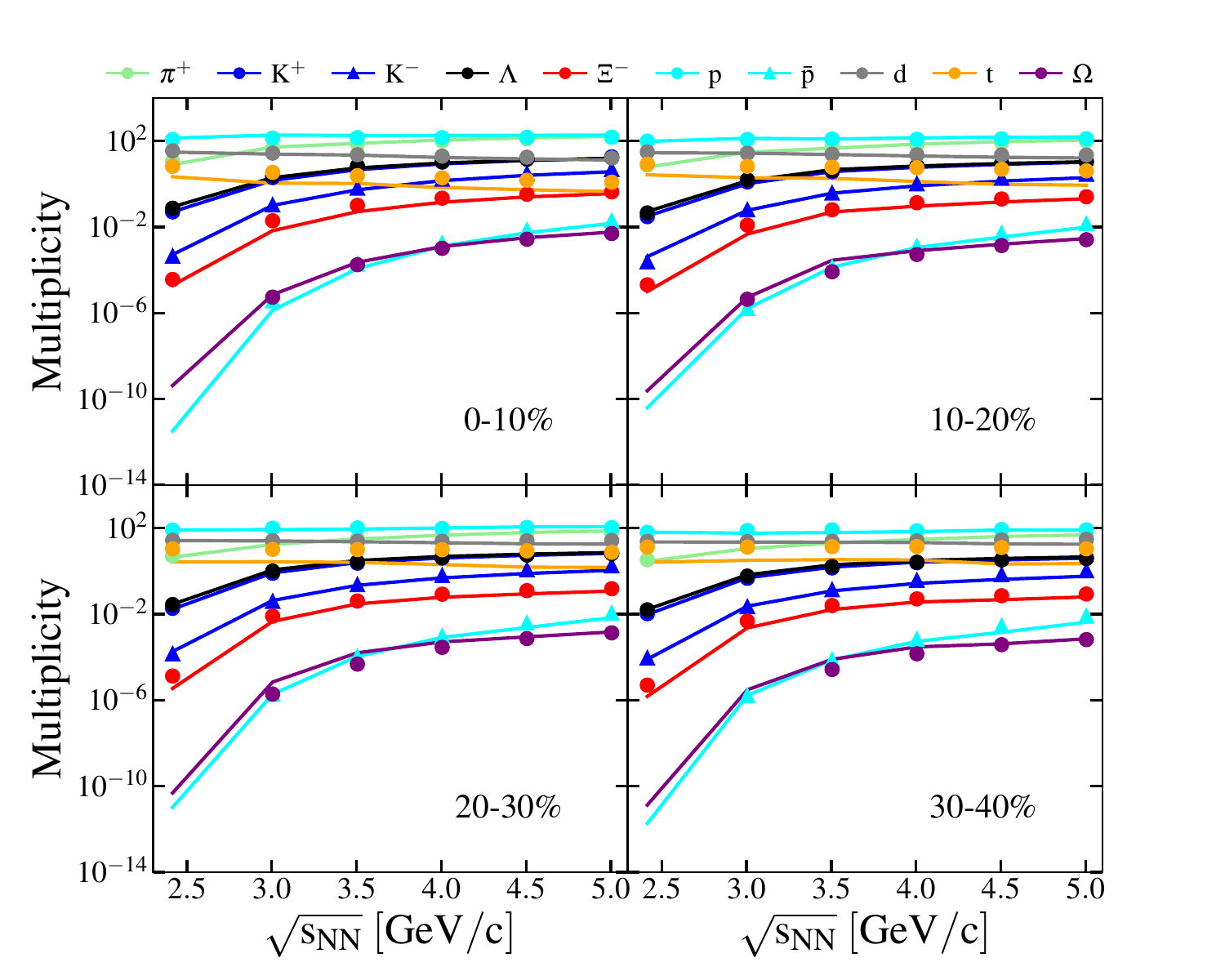}
        \caption{Input (symbols) and fitted (lines) light hadron multiplicities from the UrQMD simulations of Au+Au collisions at different centrality selections. For the fits hadron multiplicities including deuterons where used. }
        \label{fig:2}
    \end{figure}

We start by demonstrating that the hadron yields from UrQMD can indeed be described by a statistical thermal model. To show this, we calculate the full set of final state hadrons and light clusters in UrQMD and compare these to the outcome of the thermal fits for Au+Au reactions at four different centralities. Figure \ref{fig:1} shows the results of the UrQMD calculation (symbols) in comparison to the best fit using Thermal-FIST (lines), fitted to hadron yields only and neglecting clusters in the fit. The description of the UrQMD multiplicities is very good. In Fig. \ref{fig:2} we show the fit results (lines), now with the inclusion of the deuteron (as the most abundant cluster) and compare again to the UrQMD results (symbols). We observe that a very good fit to the data is also possible in this case. However, differences show up in the yields of light clusters, which is expected. This difference will later define the systematic error for charmed nuclei. In the following, we will always compare the fit without nuclei and the fit including nuclei in central reactions.

\begin{figure}[t]
        \centering
        \includegraphics[width=0.5\textwidth]{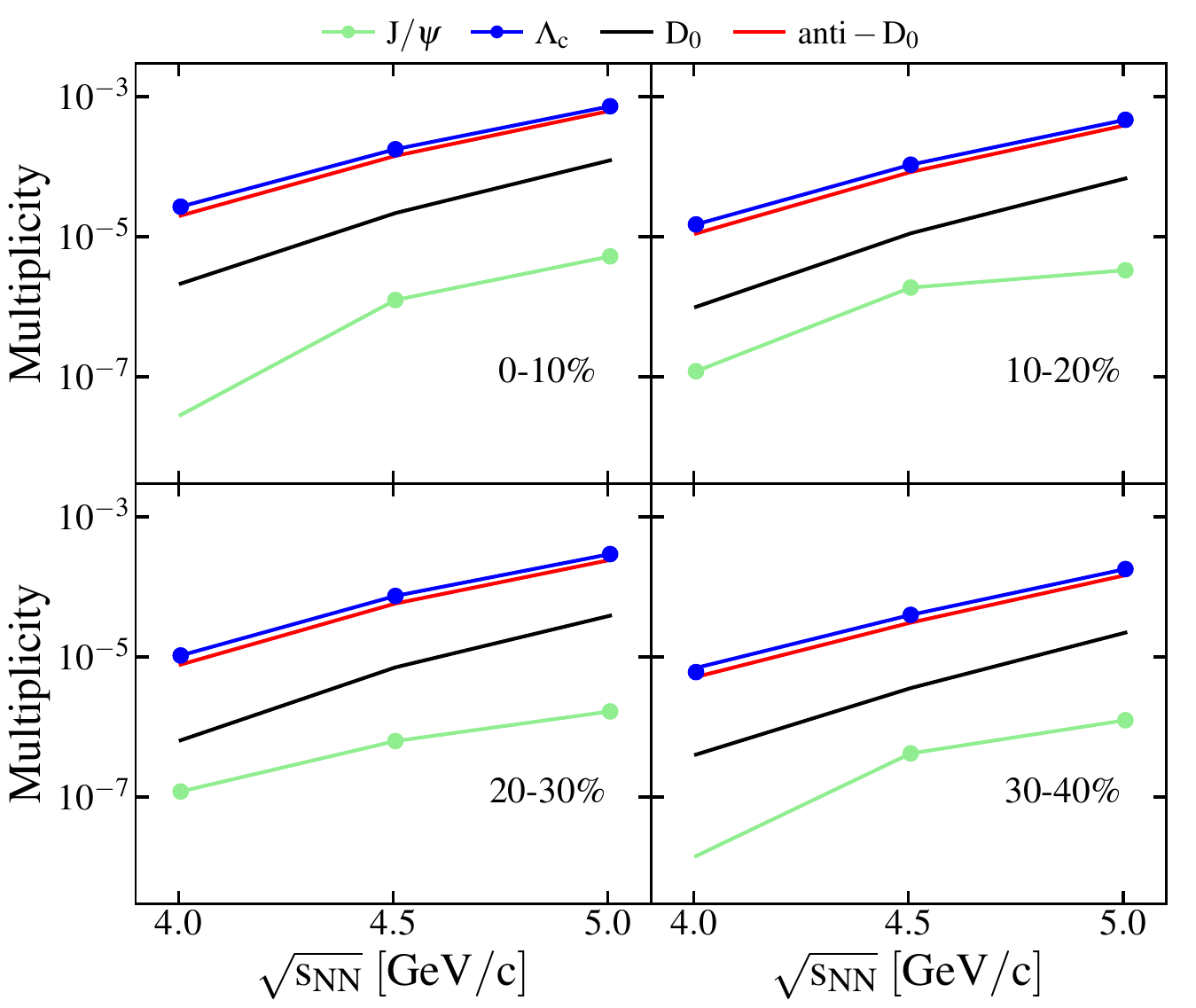}
        \caption{Input (symbols) and predicted charm hadron multiplicities in Au+Au collisions at different centralities as function of the beam energy. }
        \label{fig:3}
    \end{figure}

As benchmark for the reproduction of the input charm yields, Fig.~\ref{fig:3} shows a comparison of the thermal fit (lines) with the input multiplicities ($\Lambda_c$ and $J/\Psi$) for charmed hadrons from UrQMD (symbols). As expected, the fit is very good, simply because two parameters are fixed with two inputs. 
The yields of $D$ and $\overline D$ mesons shown are predicted and show a sizable splitting, as a consequence of a high $\mu_B$ environment.

\begin{figure}[t]
        \centering
        \includegraphics[width=0.5\textwidth]{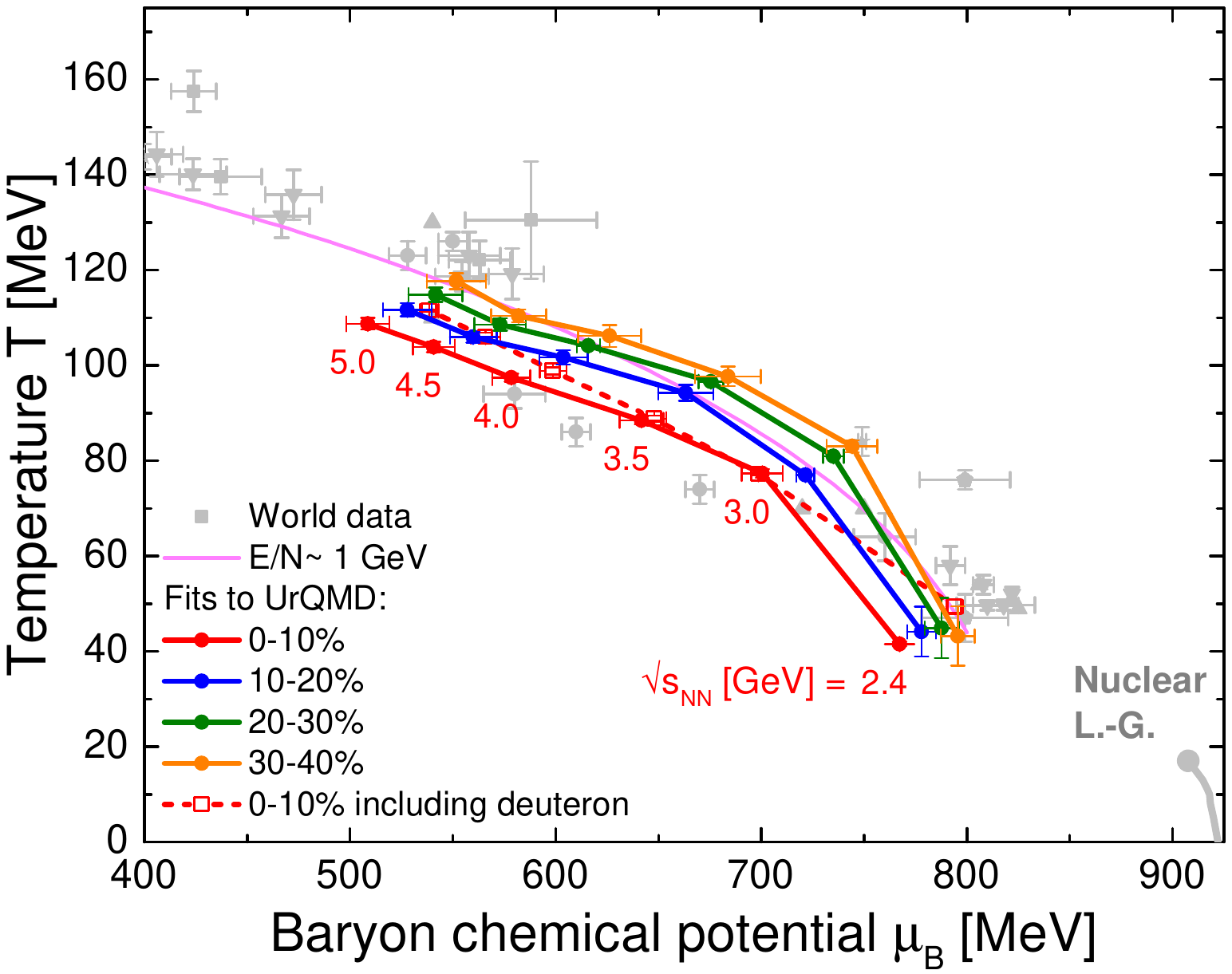}
        \caption{Freeze out points from fits to UrQMD, in the ${T-\mu_B}$ phase diagram. Different centralities are shown as filled symbols with solid lines compared to world data (grey symbols) \cite{Cleymans:1998fq,Becattini:2000jw,Cleymans:2005xv,Andronic:2005yp,HADES:2010wua,Lorenz:2014eja,Becattini:2016xct,Lysenko:2024hqp} and a parametrization of the freeze out line from \cite{Lysenko:2024hqp}. The dashed line shows the fit for central collisions when the deuteron is included in the list of fitted particles.}
        \label{fig:4}
    \end{figure}

Before diving deeper into the predictions, we want to briefly discuss the resulting freeze out line from our inputs. In Fig.\ref{fig:4}, the chemical freeze-out lines for the four different centralities (from bottom to top the centralities go from central to peripheral) are shown as colored lines with symbols (indicating the different beam energies used). These are compared to a global fit to different experiments (mostly central reactions) shown as a magenta line \cite{Andronic:2005yp}. We observe two things:

\begin{enumerate}
\item The temperature and chemical potential increases as we move from central to more peripheral collisions. This has also been observed in experiments at higher beam energies (see e.g. \cite{Becattini:2014hla}).
\item As expected and shown earlier, including the light nuclei will increase the freeze out temperature and chemical potentials leading to slightly different predictions. This is demonstrated by the red dashed line which corresponds to the fit to central (0-10$\%$) UrQMD collisions including the light nuclei in the fit (similar results have been discussed for instance in~\cite{Blaschke:2024jqd}).
\end{enumerate}

\section{Statistical model predictions for (exotic) charm states}

\begin{table}[b]
        \centering
        \scalebox{1}{ 
        \begin{tabular}{|l|c|c|c|c|c|c|c|}
        \hline
        Particle & Mass[GeV] & Q & B & S & C  & $|C|$ & $ d_J $ \\ 
        \hline\hline
       $\{\mathrm{p n} \Lambda_c\}$ & 4.163 & 2 & 3 & 0 & 1 & 1 & 2 \\
        $\{\mathrm{n} \Lambda_c \}$ & 3.225 & 1 & 2 & 0 & 1 & 1 & 3 \\
        $\{\mathrm{pnn} \Lambda_c \}$ & 5.102 & 2 & 4 & 0 & 1 & 1 & 1 \\
        $\{\alpha D^-\}$ & 5.596 & 1 & 4 & 0 & -1 & 1 & 3 \\
        X(3872) & 3.872 & 0 & 0 & 0 & 0 & 2 & 3 \\
        $D_s^+$ & 1.968 & 1 & 0 & 1 & 1 & 1 & 1 \\
        $\eta_c(1S)$ & 2.984 & 0 & 0 & 0 & 0 & 2 & 1 \\
        $\chi_{c0}(1P)$ & 3.414 & 0 & 0 & 0 & 0 & 2 & 1 \\
        $\chi_{c1}(1P)$ & 3.510 & 0 & 0 & 0 & 0 & 2 & 3 \\
        $\Sigma_c^{++}$ & 2.454 & 2 & 1 & 0 & 1 & 1 & 2\\
        $\Sigma_c^{+}$ & 2.453 & 1 & 1 & 0 & 1 & 1 & 2 \\
        \hline
    \end{tabular}
}
    \caption{List of particles and nuclei with masses, spin degeneracy and conserved quantum numbers investigated in this study.\label{tab:hadlist}}
\end{table}
    
Finally, we can move on to predictions of (exotic) charmed hadrons and charmed nuclei states. The list of hadrons and nuclei we have selected for the current study is shown in Table \ref{tab:hadlist} together with their mass and quantum numbers (the spin degeneracy is given as $d_J$). The first states, denoted with brackets $\left\{  \right\}$, are the charmed nuclei. We assume that the binding energies of charmed nuclei are negligible relative to their mass.

    \begin{figure}[t]
        \centering
        \includegraphics[width=0.5\textwidth]{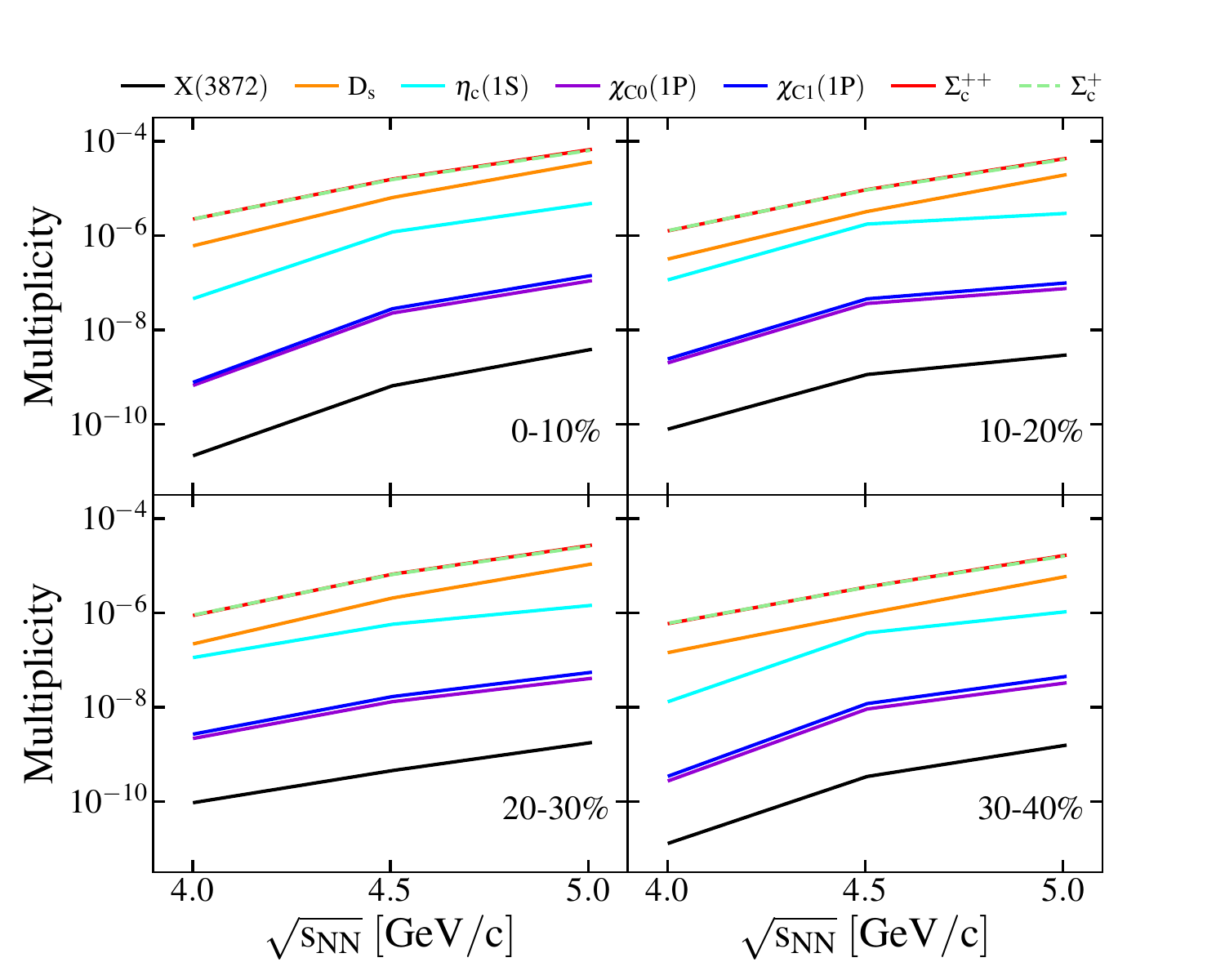}
        \caption{Predicted (lines) exotic charmed hadron multiplicities in Au+Au collisions at different centralities as function of the beam energy. }
        \label{fig:5}
    \end{figure}

We begin our discussion with Fig. \ref{fig:5}, which summarizes our predictions for the multiplicities of $X(3872)$, $D_s^+$, $\eta_c(1S)$, $\chi_{c0}(1P)$, $\chi_{c1}(1P)$, $\Sigma_c^{++}$, and  $\Sigma_c^{+}$ in central Au+Au reactions in the FAIR energy range from $\sqrt{s_{NN}}=4-5$~GeV. As expected, the most abundant charmed hadrons (after the $\Lambda_c$) are the $\Sigma_c$ states as they are the lightest states and couple to the high baryon density. They are followed by the $D_s$ state and the $\eta_c$. For the highest energy, multiplicites between $10^{-6}-10^{-4}$ are predicted per event. The next group, with multiplicites on the order of $10^{-7}$, is given by the $\chi$ states followed by the $X$ state with an estimated multiplicity around $10^{-9}$ per event. Thus, even studies of these exotic charm states seem feasible for the CBM experiment.

\begin{figure}[t]
        \centering
        \includegraphics[width=0.5\textwidth]{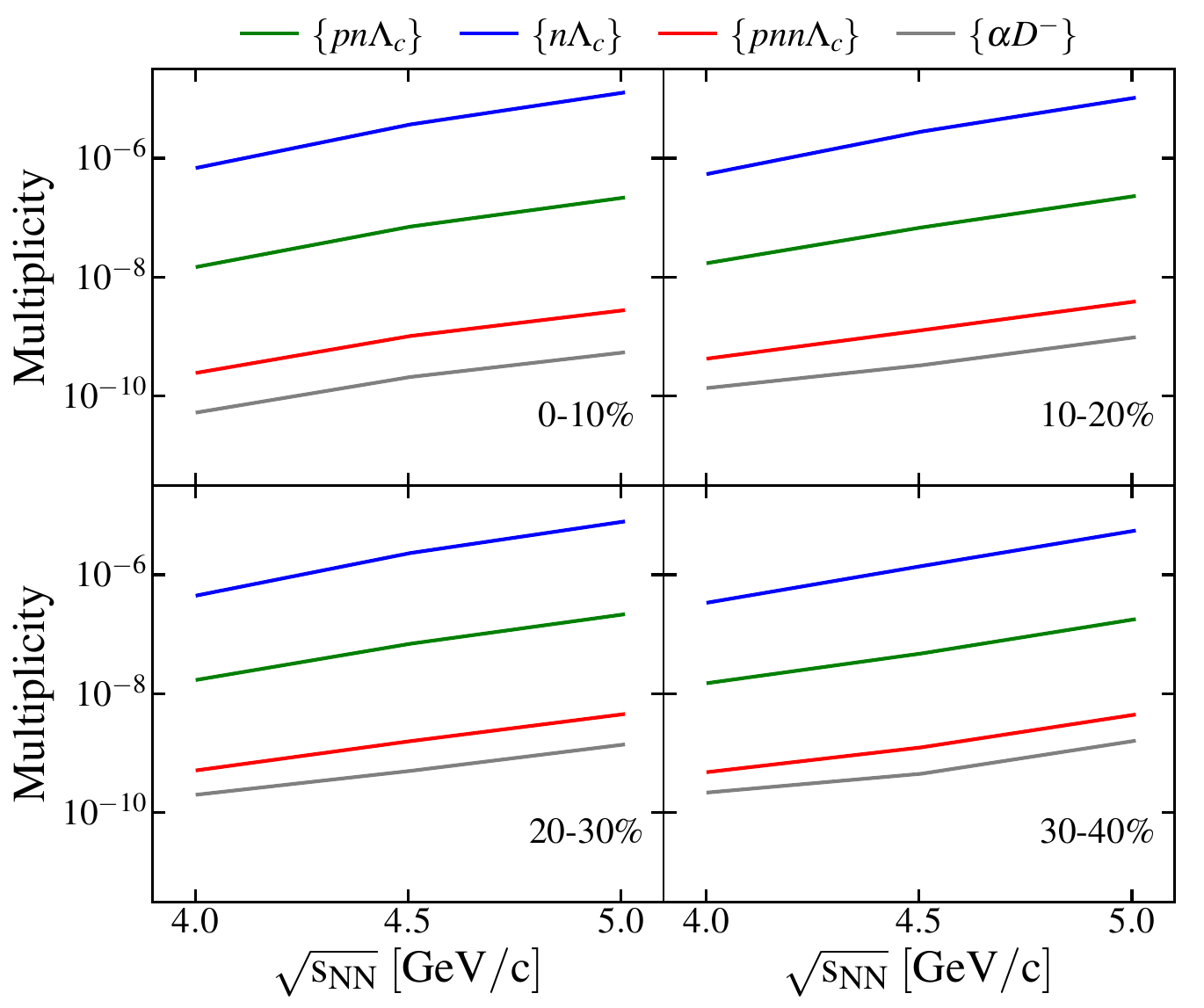}
        \caption{Predicted (lines) charmed nuclei multiplicities in Au+Au collisions at different centralities as function of the beam energy. }
        \label{fig:6}
    \end{figure}

Next, we turn to the speculated charmed nuclei and investigate the yields of $\{\mathrm{p n} \Lambda_c\}$, $\{\mathrm{n} \Lambda_c \}$,  $\{\mathrm{pnn} \Lambda_c \}$, and $\{\alpha D^-\}$. In Fig. \ref{fig:6}, these are shown as a function of energy and centrality. For the lightest charm nucleus ($\{\mathrm{n} \Lambda_c \}$), we obtain multiplicities on the order of  $10^{-7}$ per event for central events at the highest CBM energy for Au beams. As expected from the study of normal light nuclei, the addition of another nucleon comes with a penalty factor of approx 50-100 \cite{Xu:1999ac} in this energy regime, which is also what we observe. All explored charmed nuclei states have multiplicities above $10^{-10}$ per event and can therefore be explored in CBM@FAIR due to the unprecedented luminosity available.

\begin{figure*}[t]
        \centering
        \includegraphics[width=0.99\textwidth]{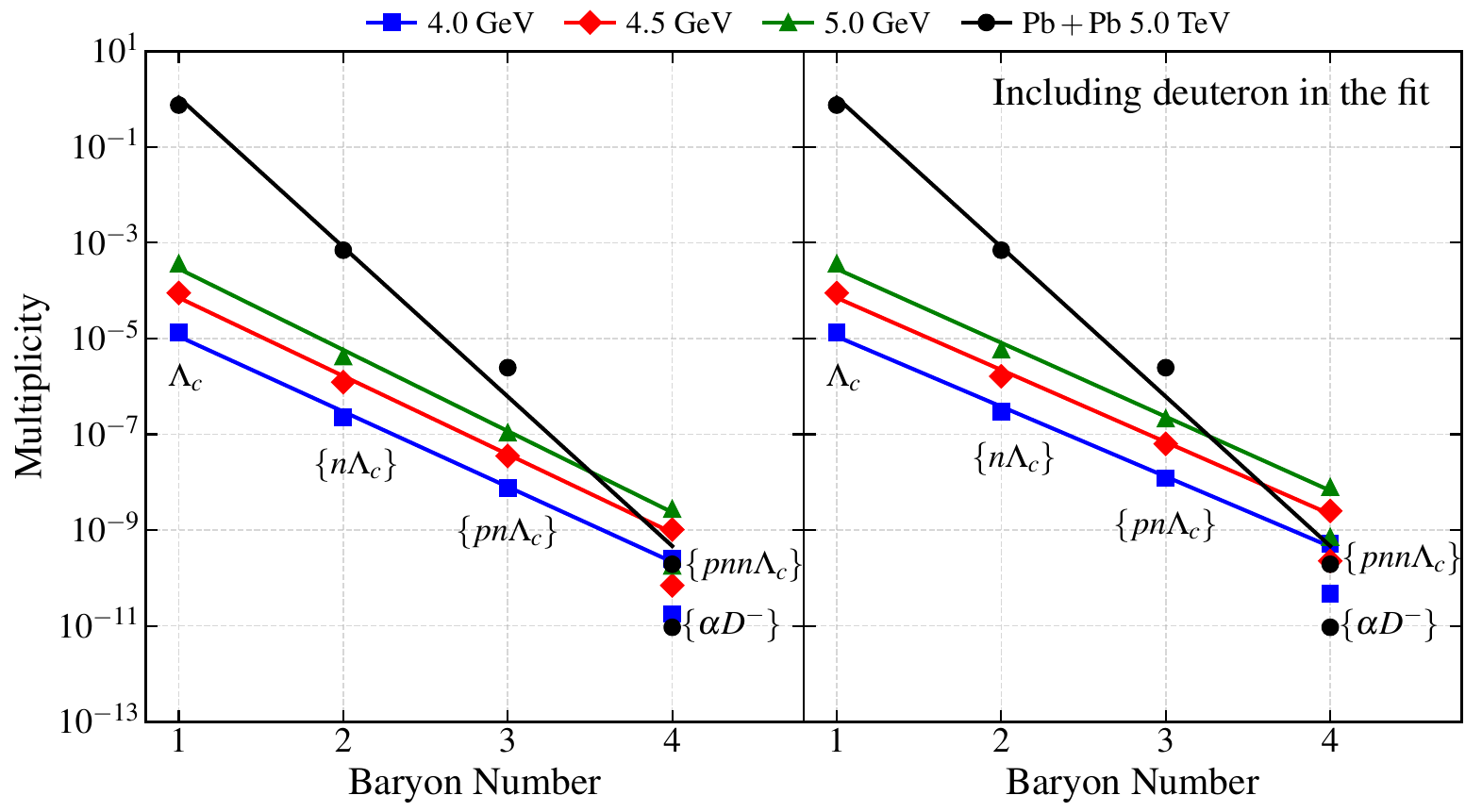}
        \caption{Multiplicities of charmed nuclei as a function of the baryon number for Au+Au collisions at 4.0 4.5 and 5.0 GeV and for Pb+Pb collisions at the LHC (5.02 TeV). The lines are exponential fits to the Thermal-FIST predictions (shown as symbols) to depict the penalty factor. The left panel shows the predictions without the deuteron in the fit, the right panel shows the predictions including the deuteron in the fit.}
        \label{fig:7}
    \end{figure*}

We now compare the discovery reach of CBM to the possibilities of ALICE at the LHC. These are two extreme scenarios for charm nuclei production: At the LHC, while the initial charm yield is high (increasing charm nuclei formation), low baryon densities and high temperatures decrease the formation probability of charmed nuclei. At FAIR energies, on the other hand, the initial charm yield is very low (decreasing charm nuclei formation), but very high baryon densities and low temperatures increase the formation probability. To explore the influence of these counter-acting effects, Fig. \ref{fig:7} shows the expected multiplicity of the charmed light nuclei as a function of the baryon number for the three different beam energies in the SIS100 beam energy range and for Pb+Pb collisions at 5.02 TeV center-of-mass energy at the LHC.
Let us clarify that the predictions for the LHC are done using measured ALICE multiplicities as input for Thermal-FIST, while in the case of FAIR, we have to rely on the model predictions as input.

The left panel of Fig. \ref{fig:7} shows the results without including the deuterons in the thermal fit and the right panel of Fig. \ref{fig:7} includes the deuterons in the fit. In general, including the deuteron in the fit leads to a small enhancement of light charmed nuclei production. However, what is even more interesting is the comparison of the predictions for SIS100 compared to the LHC. While for very light charmed nuclei ($A\leq3$) the production is suppressed compared to the LHC by up to 3 orders of magnitude, this suppression is completely compensated by the increased baryon density for charmed nuclei with $A\ge4$, resulting in an almost equal probability to produce heavier charmed nuclei.

In the following Table \ref{tab:mult} we have summarized the predicted multiplicities, integrated over all phase space, for central collisions of Au+Au (at $\sqrt{s_{NN}}= 5$ GeV) and Pb+Pb (at $\sqrt{s_{NN}}= 5.02$ TeV/c).

\section{Summary and Discussion}

We have presented first predictions for the production of exotic charmed hadrons and light charmed nuclei in heavy ion collisions at the upcoming CBM experiment. We further demonstrated the potential to observe charmed nuclei at the ALICE experiment which is currently taking data at CERN. The predictions were made by employing the Thermal-FIST model which assumes statistical hadronization and using either UrQMD model input (in the case of FAIR) or measured light flavor and charm multiplicities (in the case of ALICE). In both cases, the production rate of even the heavier {$A=4 \  (\{\mathrm{pnn} \Lambda_c \}$)} charmed nuclei is on the order of $10^{-9}$ (and for the molecular state  $\{\alpha D^-\}$ on the order of $10^{-11}$), while for lighter charm clusters, multiplicities above $10^{-8}$ ($A=3$) are reached. Given a planned collision rate of $10^7$Hz at CBM, this yields approx 1000 $\{\mathrm{pnn} \Lambda_c \}$ nuclei per day. Regarding the production of light charmed nuclei, we find that the reduced production rate of the $\Lambda_c$ at the SIS100 can be balanced by the higher baryon density which leads to almost identical production rates at the LHC and SIS100 for charmed nuclei with $A=4$.

\label{app:1}
\begin{table}[b]
        \centering
        \begin{tabular}{|l|c|c|}
        \hline
        System  & Au+Au & Pb+Pb\\ 
        \hline\hline
        $\sqrt{s_{NN}}$ &  $5$ GeV  & $5.02$ TeV  \\
        \hline\hline        
        $\{\mathrm{p n} \Lambda_c\}$ & $2.15 \times 10^{-7}$ & $4.91\times10^{-6}$ \\
        $\{\mathrm{n} \Lambda_c \}$ & $1.25 \times 10^{-5}$  & 0.002\\
        $\{\mathrm{pnn} \Lambda_c \}$ & $2.76 \times10^{-9}$ & $1.93\times10^{-10}$\\
        $\{\alpha D^-\}$ & $5.40\times10^{-10}$ & $2.8\times10^{-11}$ \\
        X(3872) & $5.71\times10^{-9}$ & 0.0015\\
        $D_s^+$ & $3.57\times10^{-5}$  & 2.91\\
        $\eta_c(1S)$ & $4.76\times10^{-6}$ & 0.132\\
        $\chi_{c0}(1P)$ & $1.09\times10^{-7}$ & 0.010\\
        $\chi_{c1}(1P)$ & $1.40\times10^{-7}$ &  0.016\\
        $\Sigma_c^{++}$ & $6.66\times10^{-5}$ & 0.159\\
        $\Sigma_c^{+}$ & $6.39\times10^{-5}$ & 0.134 \\
        \hline
    \end{tabular}
    \caption{Predicted particle multiplicities for central collisions of Au+Au (at $\sqrt{s_{NN}}= 5 \ \mathrm{GeV/c}$) and Pb+Pb (at ${\sqrt{s_{NN}}= 5.02 \ \mathrm{TeV/c}}$).}
    \label{tab:mult}
\end{table}

We have further predicted the multiplicities of the exotic charmed mesons (most of them currently under investigation at BESIII) $X(3872)$,  $\chi_{c0}(1P)$, and $\chi_{c1}(1P)$, and have demonstrated that CBM also has the capability to explore these states with a production rate of 1 per second for $\chi_{c0}(1P)$ and $\chi_{c1}(1P)$ and a rate of 1 per minute in the case of $X(3872)$.

These novel predictions can guide the planned experimental programs at CBM and ALICE to make feasibility studies on the detectability of the discussed exotic nuclear states. The study of the charmed exotic and nuclear states discussed in this work has the potential to shed new light into the physics of the strong interaction.

\begin{acknowledgments}
The computational resources for this project were provided by the GSI green cube.
TC thanks GSI for hosting her during the 2025 Summer Student program. CH, AL and YY acknowledge support from the NSRF via the Program Management Unit for Human Resources \& Institutional Development, Research and Innovation [grant number B39G680010]. 
BD acknowledges support from Bundesministerium f\"{u}r Forschung, Technologie und Raumfahrt through ErUM-FSP T01, F\"{o}rderkennzeichen 05P21RFCA1. 
VV was supported by the U.S. Department of Energy, 
Office of Science, Office of Nuclear Physics, Early Career Research Program under Award Number DE-SC0026065. TR gratefully acknowledges support from The Branco Weiss Fellowship - Society in Science, administered by the ETH Z\"urich.
\end{acknowledgments}

\end{document}